\documentclass[pra,superscriptaddress,showpacs,twocolumn,10pt]{revtex4}

\usepackage{amssymb}
\usepackage{amsmath}
\usepackage{graphicx}

\newtheorem{definition}{Definition}
\newtheorem{lemma}{Lemma}
\newtheorem{theorem}{Theorem}

\begin{document}

\title{Multipartite unlockable bound entanglement in the stabilizer formalism}
\author{Guoming Wang}
\email{wgm00@mails.tsinghua.edu.cn}
\author{Mingsheng Ying}
\email{yingmsh@tsinghua.edu.cn}

\affiliation{State Key Laboratory of Intelligent Technology and
Systems, Department of Computer Science and Technology, Tsinghua
University, Beijing, China, 100084}

\date{\today}

\begin{abstract}
We find an interesting relationship between multipartite bound
entangled states and the stabilizer formalism. We prove that if a
set of commuting operators from the generalized Pauli group on $n$
qudits satisfy certain constraints, then the maximally mixed state
over the subspace stabilized by them is an unlockable bound
entangled state. Moreover, the properties of this state, such as
symmetry under permutations of parties, undistillability and
unlockability, can be easily explained from the stabilizer
formalism without tedious calculation. In particular, the
four-qubit Smolin state [J. Smolin, Phys. Rev. A \textbf{63},
032306 (2001)] and its recent generalization to even number of
qubits [S. Bandyopadhyay \textit{et al.}, Phys. Rev. A
\textbf{71}, 062317 (2005); R. Augusiak \textit{et al.}, Phys.
Rev. A \textbf{73}, 012318 (2006)] can be viewed as special
examples of our results. Finally, we extend our results to
arbitrary multipartite systems in which the dimensions of all
parties may be different.
\end{abstract}


\maketitle

\section{Introduction}

As a peculiar phenomenon of quantum mechanics and a valuable
resource for quantum information processing such as quantum
computation \cite{S94}, quantum cryptography \cite{BB84E91},
quantum teleportation \cite{BB93} and superdense coding
\cite{BW92}, entanglement has been extensively studied during the
past years. One of the central problems about it is entanglement
distillation \cite{BB96BB96}, which is the procedure of extracting
pure entangled states from many identical copies of a mixed
entangled states by means of local operation and classical
communication(LOCC). A surprising discovery in this area is that
there exist mixed entangled states from which no pure entanglement
can be distilled out, and these states are called bound entangled
states \cite{HH98}. Much effort has been devoted to the
characterization and detection of bound entanglement
\cite{HH98,P96,H03,HH99,BD99,SS01,M061,DC001}. Moreover,
various properties and applications of bound entanglement have
been found, including its irreversibility under LOCC manipulation
\cite{YH05}, its capability of assisting the LOCC transformation
of other entangled states \cite{I04} and distilling out
classical secret bits \cite{HH05}, its violation of Bell
inequalities \cite{D01,A02,AH062}, and so on \cite{MV01,M06}.

The distillability of multipartite entangled states, however, is
much more complicated than that of bipartite entangled states. In
the most natural case, we simply say that a multipartite entangled
state is bound entangled if no pure entanglement can be distilled
between any two parties by LOCC when all the parties remain
spatially separated from each other. However, a multipartite bound
entangled state may be `unlocked' or `activated' in the following
sense: if we divide all the parties into several groups, and let
each group join together and perform collective quantum operations
(or an equivalent way is to let them share {\textit{a priori}}
singlets, since they can use them to teleport their respective
particles to a common party via quantum teleportation), then pure
entanglement may be distilled between some two different groups.
If so, this state is called an unlockable or activable bound
entangled state.

There are two famous classes of multipartite unlockable bound
entangled states that have been proposed. The first class includes a four-qubit
state called the Smolin state \cite{S01} and its recent
generalization to even number of qubits \cite{BC05,AH06}. These
states have been applied in remote information concentration
\cite{MV01}, quantum secret sharing \cite{AH06R}, and reducing
communication complexity \cite{AH06R,BZ02,BZ04}. Shor {\textit{et
al.}} also utilized the Smolin state to demonstrate a fascinating
effect named `superactivation' of bound entanglement
\cite{SS03,BR05}. In addition, in \cite{BC05} Bandyopadhyay
\textit{et al.} found that the Hilbert space of even number ($\ge
4$) of qubits can always be decomposed as a direct sum of four
orthogonal subspaces such that the normalized projectors onto the
subspaces are activable bound entangled states. The other class,
presented by D{\"{u}}r {\textit{et al.}} \cite{DC99,DC00}, has
been used to demonstrate numerous possible ways in which bound
entangled states can be activated. Besides, the relation between
multipartite distillability and Bell inequalities was also studied
in \cite{D01,A02,AS02,M061}. Despite these progresses achieved,
the general structure of multipartite unlockable bound
entanglement still remains elusive.

The stabilizer formalism \cite{G96,G97}, on the other hand, has
also played a significant role in quantum information science,
especially in quantum error correction codes \cite{S95,S96} and
cluster state quantum computation \cite{RB01}. Its essential idea
is to describe the quantum state by a set of stabilizing operators
rather than the state vector. This formalism provides a very
compact and effective way to describe and understand a lot of
phenomena in quantum information.

In this paper we link the two seemingly irrelevant areas and find
an interesting relationship between them. In specific, we prove
that if a set of commuting operators from the generalized Pauli
group on $n$ qudits satisfy certain constraints, then the
maximally mixed state over the subspace stabilized by them is an
unlockable bound entangled state, and its properties can be easily
explained from the stabilizer formalism. In particular, the Smolin
state and its generalization are reinterpreted as one special case
of our results. Furthermore, our results can also be extended to
arbitrary multipartite systems in which the dimensions of all
parties may be different.

This paper is organized as follows. In Sec. II we first briefly
recall some facts about the generalized Pauli group and the
stabilizer formalism, and then propose our main results. In Sec.
III we analyze a series of examples by using our theorems. In Sev.
IV, we extend our results to arbitrary multipartite systems.
Finally, Sec. V summarizes our results.

\section{Construction of multipartite unlockable bound entangled states}

\subsection{The generalized Pauli group and stabilizer formalism}

In this section we review some basic facts about the generalized
Pauli group and the corresponding stabilizer formalism in the
general high-dimensional case. Similar topics have also been
explored in \cite{G98,NB02,Y02,HD05}.

Consider a $d$-dimensional Hilbert space. Define \begin{equation} \begin{array}{l}
X_{(d)}=\sum\limits_{j=0}^{d-1}{|j \oplus 1\rangle\langle j|},\\
Z_{(d)}=\sum\limits_{j=0}^{d-1}{\omega^{j}|j\rangle\langle j|},
\end{array} \label{equ:pauli} \end{equation} where $\omega=e^{i\frac{2\pi}{d}}$ is
the $d$-th root of unity over the complex field and the `$\oplus$'
sign denotes addition modulo $d$. Then the matrices
$\{\sigma_{i,j}=X_{(d)}^iZ_{(d)}^j:i,j= 0,1,\dots,d-1\}$ are
considered as the generalized Pauli matrices over the
$d$-dimensional space, and they have the following commutation
relation \begin{equation}
\sigma_{i,j}\sigma_{m,n}=\omega^{jm-in}\sigma_{m,n}\sigma_{i,j}.
\label{equ:commute} \end{equation} It can be checked that when $d$
is odd, $\sigma_{i,j}$ always have eigenvalues $\{1, \omega^c,
\omega^{2c}, \dots, \omega^{d-c}\}$ for some $c|d$ (i.e. $c$ is a
factor of $d$); but when $d$ is even, the eigenvalues of
$\sigma_{i,j}$ may be either of the above form or $\{\omega^{1/2},
\omega^{c+1/2}, \omega^{2c+1/2}, \dots, \omega^{d-c+1/2}\}$ for
some $c|d$.

The generalized Pauli group on $n$ qudits $G_n$ is generated under
multiplication by the Pauli matrices acting on each qudit,
together with the phase factor $\gamma=\sqrt{\omega}$, i.e. \begin{equation}
\begin{array}{l}
G_n=\{\gamma^{a}\sigma_{i_1,j_1}\otimes\sigma_{i_2,j_2}\otimes\dots\otimes\sigma_{i_n,j_n}:
0 \le a \le 2d-1,\\
0 \le i_1,j_1,i_2,j_2,\dots,i_n,j_n \le d-1\}. \label{equ:paulig}
\end{array} \end{equation} Actually, when $d$ is odd, the introduction of $\gamma$ is
unnecessary and it can be replaced by $\omega$ (For a detailed
discussion about this, one can see \cite{HD05}). However, this
will not affect our results since in the following we consider
only elements in $G'_n =\{\bigotimes_{k=1}^{n}{\sigma_{i_k,j_k}}$:
$\forall k=1,2,\dots,n, i_k=0$ or $j_k=0\}$ $\subset G_n$. For any
element $g \in G'_n$ it has eigenvalues
$\{1,\omega^{c},\omega^{2c},\dots,\omega^{d-c}\}$ for some $c|d$.

Suppose we choose commuting operators $g_1,g_2,\dots,g_k$ from
$G'_n$. Let $S=\langle g_1,g_2,\dots,g_k \rangle$ denote the
Abelian subgroup generated by them. A state $|\psi\rangle$ is said
to be stabilized by $S$, or $S$ is the stabilizer of
$|\psi\rangle$, if $g_i|\psi\rangle= |\psi\rangle, \forall
i=1,2,\dots,k$. All the states stabilized by $S$ constitute a
subspace denoted by $V_S$. With the fact that
$\sum_{i=0}^{d-1}{\omega^{ci}}=0,\forall c=1,2,\dots,d-1$, one can
verify that the projection operator onto $V_S$ is \begin{equation}
P_S=\prod_{i=1}^{k}{\frac{(I+g_i+g_i^2+\dots+g_i^{d-1})}{d}}, \end{equation}
and the maximally mixed state over $V_S$ is $\rho_S=P_S/tr(P_S)$.
In particular, if there is a unique pure state stabilized by $S$,
i.e. $dim(V_S)=1$, $g_1,g_2,\dots,g_k$ are called a
\textit{complete} set of stabilizer generators and $S$ is called a
\textit{complete} stabilizer.

In practice we are often interested in the stabilized subspace
$V_S$, which is the subspace spanned by the simultaneous eigenstates
of the operators $\{g_1,g_2,\dots,g_k\}$ with the eigenvalues $\{1,1,\dots,1\}$.
But in general we can also consider the subspaces spanned by
the simultaneous eigenstates of $\{g_1,g_2,\dots,g_k\}$ corresponding to their
other eigenvalues $\{\lambda_1,\lambda_2,\dots,\lambda_k\}$, where $\lambda_i$
can be an arbitrary eigenvalue of $g_i$. All these subspaces have the same dimensions
and form an orthogonal decomposition of the whole space.
In particular, when $\{g_1,g_2,\dots,g_k\}$ are a complete set
of stabilizer generators, each of these subspaces is one-dimensional.

\subsection{Main results}

In the following, we define a partition of $\{1,2,\dots,n\}$ to be
a set of its proper subsets $\{T_1,T_2,\dots,T_m\}$ such that $T_i
\cap T_j=\emptyset, \forall i \neq j$ and $\cup_{i=1}^{m}
T_i=\{1,2,\dots,n\}$, and use $|T_i|$ to denote the number of
elements in $T_i$. An $n$-qudit state $\rho^{12\dots n}$ is said
to be separable with respect to a partition
$\{T_1,T_2,\dots,T_m\}$ if it can be written as \begin{equation} \rho^{12\dots
n}=\sum\limits_{k}{p^{}_k\rho^{(1)}_{k}\otimes\rho^{(2)}_{k}\otimes\dots\otimes\rho^{(m)}_k}
\end{equation} where $\sum_{k}{p_k}=1$, $p_k>0$ and $\rho^{(i)}_{k}$ is a
density operator of the subsystem $T_i$.

In order to conveniently describe our results, we introduce the
following definitions.

\begin{definition}
Suppose $g=\bigotimes_{k=1}^{n}{\sigma_{i_k,j_k}} \in G'_n$. Then
the restriction of $g$ on a subset $T \subset \{1,2,\dots,n\}$ is
defined as $g^{(T)}=\bigotimes_{k \in T}{\sigma_{i_k,j_k}}$.
\end{definition}
\begin{definition}
Two operators $g,h \in G'_n$ are said to commute locally with
respect to a partition $\{T_1,T_2\dots,T_m\}$ of $\{1,2,\dots,n\}$
if $g^{(T_\alpha)}h^{(T_\alpha)}=h^{(T_\alpha)}g^{(T_\alpha)},
\forall \alpha =1,2,\dots,m$.
\end{definition}
\begin{definition}
Suppose $g_1,g_2,\dots,g_k$ are commuting elements in $G'_n$.
$S=\langle g_1,g_2,\dots,g_k \rangle$ is said to be separable with
respect to a partition $\{T_1,T_2,\dots,T_m\}$ of
$\{1,2,\dots,n\}$ if $g_1,g_2\dots,g_k$ commute locally with
respect to this partition. Otherwise, if such a partition does not
exist, $S$ is said to be inseparable.
\end{definition}
Note that in the third definition, the separability of a
stabilizer with respect to any partition does not depend on the
choice of its generators, so it is well-defined.

The following lemma establishes a connection between the
separability of a stabilizer $S$ and the separability of the
maximally mixed state over the stabilized subspace $V_S$:

\begin{lemma}
Suppose $g_1,g_2,\dots,g_k$ are commuting elements in $G'_n$.
$S=\langle g_1,g_2\dots,g_k \rangle$ is separable with respect to
a partition $\{T_1,T_2\dots,T_m\}$ of $\{1,2,\dots,n\}$ if and
only if the maximally mixed state $\rho_S$ over the stabilized
subspace $V_S$ is separable with respect to the same partition. So
if $S$ is inseparable, then $\rho_S$ is a genuine $n$-qudit
entangled state.
\end{lemma}

{\it Proof:}``$\Longrightarrow$": Suppose $S=\langle
g_1,g_2,\dots,g_k\rangle$ is separable with respect to a partition
$\{T_1,T_2,\dots,T_m\}$. Then for $\forall \alpha =1,2,\dots,m$,
the operators
$g_1^{(T_\alpha)},g_2^{(T_\alpha)},\dots,g_k^{(T_\alpha)}$ are
mutually commutative and thus can be simultaneously diagonalized.
Suppose $\{|\psi^{(\alpha)}_{\beta_\alpha}\rangle:\beta_\alpha=
1,2,\dots,d^{|T_{\alpha}|}\}$ are their simultaneous eigenstates
corresponding to the eigenvalue
$\lambda^{\alpha}_{\beta_\alpha,j}$ for each $j=1,2,\dots,k$. Then
it is obvious that the $n$-qudit states
$|\psi_{\beta_1,\beta_2,\dots,\beta_m}\rangle\equiv\bigotimes_{\alpha=1}^{m}|\psi^{(\alpha)}_{\beta_\alpha}\rangle$
are the simultaneous eigenstates of
$\{g_j=\bigotimes_{\alpha=1}^{m}{g^{(T_{\alpha})}_{j}}\}$ with the
eigenvalue $\Pi_{\alpha=1}^{m}\lambda^{\alpha}_{\beta_\alpha,j}$
for each $j =1,2,\dots,k$. They also form an orthonormal basis of
the $n$-qudit space. In particular, let
$P=\{(\beta_1,\beta_2,\dots,\beta_m):
\Pi_{\alpha=1}^{m}\lambda^{\alpha}_{\beta_\alpha,j}=1, \forall j
=1,2,\dots,k\}$. Then we have \begin{equation}
\rho_S=\frac{1}{|P|}\sum\limits_{(\beta_1,\beta_2,\dots,\beta_m)
\in P}
{\bigotimes_{\alpha=1}^{m}{|\psi^{(\alpha)}_{\beta_\alpha}\rangle\langle\psi^{(\alpha)}_{\beta_\alpha}|}},
\label{equ:rhopsi} \end{equation} which implies that $\rho_S$ is
separable with respect to the partition $\{T_1,T_2,\dots,T_m\}$.

``$\Longleftarrow$": Suppose $\rho_S$ is separable with respect to
the partition $\{T_1,T_2,\dots,T_m\}$. Then there exists a state
$|\psi\rangle \in V_S$ such that $|\psi\rangle$ can be written as
$|\psi\rangle=\bigotimes_{\alpha=1}^{m}|\psi^{(\alpha)}\rangle$,
where $|\psi^{(\alpha)}\rangle$ is a state of the subsystem
$T_\alpha$. Since $|\psi\rangle$ is stabilized by $S$, we have
$|\psi\rangle=g_j|\psi\rangle=\bigotimes_{\alpha=1}^{m}g^{(T_\alpha)}_j|\psi^{(\alpha)}\rangle,
\forall j=1,2,\dots,k$, which means that $|\psi^{(\alpha)}\rangle$
should be a simultaneous eigenstate of
$g^{(T_\alpha)}_1,g^{(T_\alpha)}_2,\dots,g^{(T_\alpha)}_k$ for
each $\alpha =1,2,\dots,m$. This is impossible if
$g^{(T_\alpha)}_1,g^{(T_\alpha)}_2,\dots,g^{(T_\alpha)}_k$ do not
commute. To see this, we prove that any two elements $g,h \in
G'_l$ for any $l$ do not have a simultaneous eigenstate if $g,h$
do not commute. From Eq.(\ref{equ:commute}) and
Eq.(\ref{equ:paulig}) one can see that $gh=\omega^{f(g,h)}hg$ for
some integer $f(g,h)$ determined by $g$ and $h$. If $g$ and $h$ do
not commute, i.e. $\omega^{f(g,h)} \ne 1$, and they share a
simultaneous eigenstate $|\psi\rangle$ which corresponds to the
eigenvalues $\lambda$, $\mu$ of $g$, $h$ respectively, then we
have
\begin{equation} \begin{array}{l}
gh|\psi\rangle=g(\mu|\psi\rangle)=\lambda\mu|\psi\rangle \\
=\omega^{f(g,h)}hg|\psi\rangle=\omega^{f(g,h)}h(\lambda|\psi\rangle)=\omega^{f(g,h)}\mu\lambda|\psi\rangle,
\end{array} \end{equation} which implies that at least one of $\lambda$ and $\mu$
must be zero. But this contradicts with the fact that any operator
in the generalized Pauli group has only nonzero eigenvalues. So
$g_1,g_2,\dots,g_k$ commute locally with respect to the partition
$\{T_1,T_2,\dots,T_m\}$ and $S=\langle g_1,g_2,\dots,g_k \rangle$
is separable with respect to this partition. \hfill $\blacksquare$

With the help of Lemma 1, we find that the distillability and
unlockability of $\rho_S$ generated by an incomplete stabilizer
$S=\langle g_1,g_2,\dots,g_k\rangle$ are determined by the
separability of $S$, as the following theorem states:

\begin{theorem}
Suppose $g_1,g_2,\dots,g_k$ are commuting elements in $G'_n$. Let
$S=\langle g_1,\dots,g_k\rangle$. If

(1)for any $i \ne j \in \{1,2,\dots,n\}$, there exits a partition
$\{Q_1,Q_2,\dots,Q_m\}$ with $i \in Q_1$, $j \in Q_2$ such that
$S$ is separable with respect to this partition.

(2)there exists a partition $\{T_1,T_2,\dots,T_m\}$ with $|T_1|>1$
such that $S$ is separable with respect to this partition and
$S^{(T_1)}=\langle g^{(T_1)}_1,g^{(T_1)}_2,\dots,g^{(T_1)}_k
\rangle$ is an inseparable and complete stabilizer on $T_1$.

Then the maximally mixed state $\rho_S$ over the stabilized
subspace $V_S$ is an unlockable bound entangled state. Moreover,
for any partition $\{T_1,T_2,\dots,T_m\}$ satisfying condition
(2), pure entanglement among the parties inside $T_1$ can be
distilled by letting the parties inside $T_2,T_3,\dots,T_m$ join
together respectively.
\end{theorem}

{\it Proof}: First, we prove that $\rho_S$ is undistillable.
Consider any two parties $i,j \in \{1,2,\dots,n\}$. By condition
(1) and Lemma 1 we can find a partition $\{Q_1,Q_2,\dots,Q_m\}$
with $i \in Q_1$ and $j \in Q_2$ such that $\rho_S$ is separable
with respect to it. So it is impossible to distill out pure
entanglement between $i$ and $j$, even between $Q_1$ and $Q_2$, by
LOCC, as long as $Q_1$ and $Q_2$ remain spatially separated.

Next, we prove that $\rho_S$ can be unlocked. Consider the
partition $\{T_1,T_2,\dots,T_m\}$ which fulfills condition (2).
Since $S$ is separable with respect to this partition, we can
repeat exactly the same argument presented in the first part of
the proof of Lemma 1 without changing any notations introduced.
Now suppose all the parties inside $T_{\alpha}$ join together and
perform the projection measurement in the basis
$\{|\psi^{(\alpha)}_{\beta_\alpha}\rangle:\beta_\alpha=1,\dots,d^{|T_{\alpha}|}\}$
for each $\alpha=2,3,\dots,m$, and obtain the outcomes
$\beta'_2,\beta'_3,\dots,\beta'_m$ respectively. Then by
Eq.(\ref{equ:rhopsi}) we have the remaining state of the subsystem
$T_1$ is
\begin{equation}
\rho^{(1)}_S=\frac{1}{|P_{\beta'_2,\beta'_3,\dots,\beta'_m}|}
\sum\limits_{\beta_1 \in
P_{\beta'_2,\beta'_3,\dots,\beta'_m}}{|\psi^{(1)}_{\beta_1}\rangle
\langle \psi^{(1)}_{\beta_1}|},
\end{equation}
where
$P_{\beta_2,\beta_3,\dots,\beta_m}=\{\beta_1:
\lambda^{1}_{\beta_1,j}={1}/{\Pi_{\alpha=2}^{m}\lambda^{\alpha}_{\beta_\alpha,j}},
\forall j =1,2,\dots,k\}$. Since $S^{(T_1)}=
\langle g^{(T_1)}_1,g^{(T_1)}_2,\dots,g^{(T_1)}_k \rangle$ is a complete
stabilizer on $T_1$, we have that $P(\beta'_2,\beta'_3,\dots,\beta'_m)$ actually
contains only one element and therefore $\rho^{(1)}_S$ is a pure state.
Moreover, because $S^{(T_1)}$ is inseparable, by Lemma 1
we know that $\rho^{(1)}_S$ is a genuine $|T_1|$-qudit
entangled state. Therefore we have obtained an activation
strategy.

\hfill $\blacksquare$

Note that by a similar argument, we can easily prove that
Lemma 1 and Theorem 1 will still hold if we replace
$\rho_S$ by a maximally mixed state over the subspace spanned by
the simultaneous eigenstates of $\{g_1,g_2,\dots,g_k\}$ corresponding to
their eigenvalues $\{\lambda_1,\lambda_2,\dots,\lambda_k\}$,
where $\lambda_i$ is an arbitrary eigenvalue of $g_i$.
Recalling that all these subspaces have the same dimensions,
we reach the following conclusion:

\begin{theorem}
Suppose $g_1,g_2,\dots,g_k$ are $k$ commuting elements in $G'_n$.
If they satisfy the condition (1) and (2) in Theorem 1, and
the subspace stabilized by them is $b$-dimensional with $b|d^n$,
then the Hilbert space of $n$ qudits can be decomposed into
$\frac{d^n}{b}$ orthogonal subspaces such
that the normalized projection operator onto each of them is an
unlockable bound entangled state.
\label{thm2}
\end{theorem}

The two theorems above provide a simple method of constructing a
wide class of unlockable bound entangled states in arbitrary
multiqudit systems. What we need to do now is to appropriately
choose several commuting operators from the generalized Pauli
group on $n$ qudits. It is worth noting that our construction
essentially utilizes the symmetry of the generalized Pauli
matrices. Consequently the constructed states also own some
inherent symmetry. With the help of Lemma $1$, the properties of
these states can be easily explained from the stabilizer
formalism, as shown in the subsequent section.

\section{Illustrations}

In this section we will analyze several concrete examples by using
our theorems. Without explicitly pointed out, the matrices $X$ and
$Z$ appearing below are $X_{(d)}$ and $Z_{(d)}$ defined by
Eq.(\ref{equ:pauli}) with the corresponding dimension $d$. We will
also use the notation $X_j$ to denote the operation $X$ acting on
the $j$th party and similarly for $Z_j$.

Example 1: Consider a $4$-qubit system. Define \begin{equation} \begin{array}{l}
g_1=X_1X_2X_3X_4,\\
g_2=Z_1Z_2Z_3Z_4.\\
\end{array} \end{equation} The maximally mixed state over the subspace stabilized by
$g_1$ and $g_2$ is \begin{equation} \rho^{(4)}\equiv\rho_{\langle g_1,g_2
\rangle}=\frac{1}{16}(I+g_1)(I+g_2) \end{equation}

Because $X \otimes X$ and $Z \otimes Z$ commute, $S=\langle
g_1,g_2 \rangle$ is separable with respect to any $2:2$ partition
of $\{1,2,3,4\}$, which assures that the condition (1) in Theorem
1 is fulfilled. Any $2:2$ partition also satisfies the condition
(2) in Theorem 1 since $S=\langle X \otimes X, Z \otimes Z\rangle$
is an inseparable and complete stabilizer on two qubits. So
$\rho^{(4)}$ is an unlockable bound entangled state and pure
entanglement can be distilled between any two parties.

Actually, this state is exactly the Smolin state which is
originally defined as
\begin{equation} \begin{array}{l}
\rho^{(4)}=\frac{1}{4}\sum\limits_{\alpha,\beta=0}^{1}|\Phi_{\alpha\beta}\rangle_{12}\langle\Phi_{\alpha\beta}|
\otimes|\Phi_{\alpha\beta}\rangle_{34}\langle\Phi_{\alpha\beta}|,\\
\end{array} \label{equ:smolin} \end{equation}
where
\begin{equation} \begin{array}{ll}
|\Phi_{00}\rangle=\frac{1}{\sqrt{2}}(|00\rangle+|11\rangle),
& |\Phi_{01}\rangle=\frac{1}{\sqrt{2}}(|00\rangle-|11\rangle),\\
|\Phi_{10}\rangle=\frac{1}{\sqrt{2}}(|01\rangle+|10\rangle),
& |\Phi_{11}\rangle=\frac{1}{\sqrt{2}}(|01\rangle-|10\rangle)\\
\end{array} \end{equation}
are the four Bell states. To see this, one only need to
realize that $|\Phi_{00}\rangle, |\Phi_{01}\rangle,
|\Phi_{10}\rangle, |\Phi_{11}\rangle$ are the simultaneous
eigenstates of $\{X\otimes X, Z \otimes Z\}$, with the eigenvalues
$\{+1,+1\}$, $\{-1,+1\}$,$\{+1,-1\}$,$\{-1,-1\}$, respectively.
Considering the $2:2$ partition $\{\{1,2\},\{3,4\}\}$, we have
$g^{(\{1,2\})}_1=X_1X_2$, $g^{(\{1,2\})}_2=Z_1Z_2$,
$g^{(\{3,4\})}_1=X_3X_4$ and $g^{(\{3,4\})}_2=Z_3Z_4$. So the four
states
$\{|\Phi_{\alpha\beta}\rangle_{12}|\Phi_{\alpha\beta}\rangle_{34}:
\alpha,\beta=0,1\}$ are the simultaneous eigenstates of
$g_1=g^{(\{1,2\})}_1\otimes g^{(\{3,4\})}_1$ and
$g_2=g^{(\{1,2\})}_2\otimes g^{(\{3,4\})}_2$ with the eigenvalues
$\{1,1\}$. Thus by Eq.(\ref{equ:rhopsi}) $\rho^{(4)}$ can be
written in the form of Eq.(\ref{equ:smolin}).

Furthermore, one can repeat the above argument by considering two
other $2:2$ partitions $\{\{1,3\},\{2,4\}\}$ and
$\{\{1,4\},\{2,3\}\}$, and can easily concludes that $\rho^{(4)}$
can also be written as \begin{equation} \begin{array}{ll}
\rho^{(4)}&=\frac{1}{4}\sum\limits_{\alpha,\beta=0}^{1}|\Phi_{\alpha\beta}\rangle_{13}\langle\Phi_{\alpha\beta}|
\otimes|\Phi_{\alpha\beta}\rangle_{24}\langle\Phi_{\alpha\beta}|\\
&=\frac{1}{4}\sum\limits_{\alpha,\beta=0}^{1}|\Phi_{\alpha\beta}\rangle_{14}\langle\Phi_{\alpha\beta}|
\otimes|\Phi_{\alpha\beta}\rangle_{23}\langle\Phi_{\alpha\beta}|,\\
\label{equ:smolin2} \end{array} \end{equation} which implies that $\rho^{(4)}$ is
invariant under arbitrary permutation of the four parties. Note
that this symmetry essentially arises from the fact that $g_1$ and
$g_2$ both act identically on the four qubits.

By Eq.(\ref{equ:smolin}) and Eq.(\ref{equ:smolin2}), $\rho^{(4)}$
is separable with respect to any $2:2$ partition, and moreover,
when any two parties get together and perform the projective
measurement in the Bell basis, if their subsystem collapses into
the state $|\Phi_{\alpha\beta}\rangle$, then the other two parties
are in the same state $|\Phi_{\alpha\beta}\rangle$.

Example 2: Consider a system of $2n(n \ge 2)$ qubits. Define \begin{equation}
\begin{array}{l}
g^{(2n)}_1=X_1X_2X_3X_4\dots X_{2n-1}X_{2n},\\
g^{(2n)}_2=Z_1Z_2Z_3Z_4\dots Z_{2n-1}Z_{2n}.\\
\end{array} \end{equation} Then the maximally mixed state over the subspace
stabilized by $g^{(2n)}_1$ and $g^{(2n)}_2$ is \begin{equation} \begin{array}{l}
\rho^{(2n)}\equiv\rho_{\langle g^{(2n)}_1,g^{(2n)}_2\rangle}
=\frac{1}{4^{n}}(I+g^{(2n)}_1)(I+g^{(2n)}_2). \end{array} \end{equation}

One can easily check that $S=\langle g^{(2n)}_1,g^{(2n)}_2
\rangle$ is separable with respect to any $2:2:\dots:2$ partition
of $\{1,2,\dots,2n\}$, which ensures the satisfaction of condition
(1) in Theorem 1. Moreover, any $2:2:\dots:2$ partition satisfies
the condition (2) in Theorem 1. So $\rho^{(2n)}$ is an unlockable
bound entangled state and pure entangled state can be distilled
between any two parties by letting the other $2n-2$ parties
pairwise group together.

Actually, $\rho^{(2n)}$ is equivalent to the generalized Smolin
state proposed in \cite{BC05} and \cite{AH06}, up to an
unimportant local Pauli operation. To see this, consider the
$(2n-2):2$ partition $\{\{1,2,\dots,2n-2\},\{2n-1,2n\}\}$. It is
observed that $g^{(2n)}_1, g^{(2n)}_2$ commute locally with
respect to this partition, and their restrictions on the subset
$\{1,2,\dots,2n-2\}$ are $g^{(2n-2)}_1, g^{(2n-2)}_2$
respectively. Let $\sigma_{00}=I_{1}$, $\sigma_{01}=Z_{1}$,
$\sigma_{10}=X_{1}$, $\sigma_{11}=Y_{1}$ be the four Pauli
operations acting on the first qubit. Then
$\sigma^{}_{\alpha\beta}\rho^{(2n-2)}\sigma^{\dagger}_{\alpha\beta}$
is actually the maximally mixed state over the subspace spanned
by the simultaneous eigenstate of $g^{(2n-2)}_1, g^{(2n-2)}_2$ with the eigenvalues
$\{(-1)^{\beta},(-1)^{\alpha}\}$. Conseqently by
Eq.(\ref{equ:rhopsi}) we have \begin{equation} \begin{array}{l}
\rho^{(2n)}=\frac{1}{4}\sum\limits_{\alpha,\beta=0}^{1}
\sigma^{}_{\alpha\beta}\rho^{(2n-2)}\sigma^{\dagger}_{\alpha\beta}
\otimes |\Phi_{\alpha\beta}\rangle\langle\Phi_{\alpha\beta}|,
\label{equ:smolin2nre} \end{array} \end{equation} which is the recursive definition
of the generalized Smolin states in \cite{BC05,AH06} up to an
local Pauli operation. Moreover, continuing this induction on $n$,
one could at last get \begin{equation} \begin{array}{l}
\rho^{(2n)}=\frac{1}{4^{n-1}}\sum\limits_{
\oplus_{i=1}^{n}{\alpha_i}=\oplus_{i=1}^{n}{\beta_i}=0}
{\bigotimes_{i=1}^{n}
|\Phi_{\alpha_i\beta_i}\rangle\langle\Phi_{\alpha_i\beta_i}|},
\label{equ:smolin2n} \end{array} \end{equation} where $\oplus$ denotes addition
modulo $2$. Noting that the two stabilizer generators $g^{(2n)}_1$
and $g^{(2n)}_2$ both act symmetrically on $2n$ qubits, one can
find that $\rho^{(2n)}$ is invariant under arbitrary permutation
of parties, which means Eq.(\ref{equ:smolin2n}) holds not only for
the partition $\{\{1,2\},\{3,4\},\dots,\{2n-1,2n\}\}$ but also for
arbitrary $2:2:\dots:2$ partition.

Now suppose any $2n-2$ parties join together pairwise and perform
the projective measurement in the Bell basis. If the $n-1$
obtained outcomes are $|\Phi_{\alpha_2,\beta_2}\rangle,
|\Phi_{\alpha_3,\beta_3}\rangle,\dots,|\Phi_{\alpha_n,\beta_n}\rangle$
respectively, then by Eq.(\ref{equ:smolin2n}) the remaining two
parties get one of the four Bell states
$|\Phi_{\alpha_1,\beta_1}\rangle$ with
$\alpha_1=\oplus_{i=2}^{n}{\alpha_i}$ and
$\beta_1=\oplus_{i=2}^{n}{\beta_i}$.

In addition, by applying Theorem 2 to $g^{(2n)}_1, g^{(2n)}_2$ we
know that the Hilbert space of $2n(\ge 2)$ qubits can be
decomposed into four orthogonal subspaces such that the normalized
projection operator onto each of them is an unlockable bound
entangled state, which was first pointed out in \cite{BC05}.

One may wonder whether there exists an analog of the Smolin state
in systems of odd number of qubits. We believe that such a state
is unlikely to exist, and even if it exists, it cannot be obtained
by our method. Because if we want the constructed state to be
symmetric under arbitrary permutation of parties, all the
stabilizer generators should act equally on each qubit. But the
tensor products of odd number of $X$'s and $Z$'s, or $X$'s and
$Y$'s, or $Y$'s and $Z$'s, do not commute. Instead they
anti-commute, e.g. $X^{\otimes 2n+1}Z^{\otimes 2n+1}=-Z^{\otimes
2n+1}X^{\otimes 2n+1}$. Therefore they cannot be simultaneously
used as stabilizer generators.

From Example 1 and 2, we can see that the properties of the Smolin
state and its generalization become so clear when they are
redefined and reinterpreted in the stabilizer formalism. However,
they are only two special instances which own the strongest
symmetry. At the cost of losing symmetry to different extents,
many more unlockable bound entangled states can be found in a
similar way.

Example 3: Consider a $9$-qubit system. Let \begin{equation} \begin{array}{l}
g_1=X_1X_2Z_3X_4X_5Z_6X_7X_8Z_9,\\
g_2=X_1Z_2X_3X_4Z_5X_6X_7Z_8X_9,\\
g_3=Z_1X_2X_3Z_4X_5X_6Z_7X_8X_9.\\
\end{array} \end{equation} The maximally mixed state over the subspace stabilized by
them is \begin{equation} \rho_{\langle g_1,g_2,g_3
\rangle}=\frac{1}{2^9}(I+g_1)(I+g_2)(I+g_3). \end{equation}

The nine qubits of this state can be classified into three groups:
$\{1,4,7\}, \{2,5,8\}$, and $\{3,6,9\}$. $g_1,g_2$ and $g_3$ all
act symmetrically on the three qubits of each group. So the state
remains invariant when exchanging any two parties inside the same
group. However, when exchange two parties that belongs to two
different groups, such as $1$ and $6$, the state will change.

Now consider two different partitions:
$\{\{1,2,3\},\{4,5,6\},\{7,8,9\}\}$ and
$\{\{1,6,8\},\{2,4,9\},\{3,5,7\}\}$. It can be verified that
$S=\langle g_1,g_2,g_3 \rangle$ is separable with respect to both
of them and this fact ensures the satisfaction of the condition
(1) in Theorem 1. Furthermore, the first partition
$\{\{1,2,3\},\{4,5,6\},\{7,8,9\}\}$ also satisfies condition (2)
in Theorem 1. Therefore, $\rho_{\langle g_1,g_2,g_3 \rangle}$ is
an unlockable bound entangled state and it can be unlocked as
follows: let the parties $4,5,6$ join together and similarly for
$7,8,9$. Then each of the two groups performs the projective
measurement in the basis of the simultaneous eigenstates of three
operators $\{X \otimes X \otimes Z, X \otimes Z \otimes X, Z
\otimes X \otimes X\}$, then depending on their measurement
outcomes a genuine three-qubit pure entangled state, which is also
a simultaneous eigenstate of the three operators, is distilled out
among the parties $1,2$ and $3$. In addition, by the symmetry of
$\rho_{\langle g_1,g_2,g_3 \rangle}$ presented above, we know that
any three parties $i \in \{1,4,7\}$, $j \in \{2,5,8\}$ and $k \in
\{3,6,9\}$ can obtain a genuine three-qubit pure entangled state
among them by appropriately grouping the other six parties.

Example 4: Consider a $7$-qutrit system, i.e. $d=3$. Let \begin{equation}
\begin{array}{l}
g_1=X^{2}_1Z_2Z^{2}_3X_4Z^{2}_5X_6Z_7,\\
g_2=Z_1X_2X^{2}_3Z_4X^{2}_5Z_6X_7.\\
\end{array} \end{equation} The maximally mixed state over the subspace stabilized by
them is \begin{equation} \rho_{\langle g_1,g_2
\rangle}=\frac{1}{3^7}(I+g_1+g^2_1)(I+g_2+g^2_2). \end{equation}

The seven qutrits of this state can be classified into four
groups: $\{1\}, \{2,7\}, \{3,5\}$ and $\{4,6\}$. $g_1,g_2$ both
act symmetrically on the qutrits of each group. So the state
remains invariant when exchanging any two parties inside the same
group. However, it will vary when exchange two parties that
belongs to two different groups, such as $1$ and $2$.

Consider two partitions: $\{\{1,2,3\},\{4,5\},\{6,7\}\}$ and
$\{\{1,4\},\{2,5,7\},\{3,6\}\}$. It can be checked that $S=\langle
g_1, g_2 \rangle$ is separable with respect to both of them, which
makes the condition (1) in Theorem 1 fulfilled. Also, the
partition $\{\{1,4\},\{2,5,7\},\{3,6\}\}$ satisfies condition (2)
in Theorem 1. So $\rho_{\langle g_1,g_2 \rangle}$ is an unlockable
bound entangled state and can be activated in the following way:
let the parties $2,5,7$ join together and similarly for $3,6$.
Then the first groups perform the projective measurement in the
basis of the simultaneous eigenstates of the operators
$\{Z_2Z^{2}_5Z_7, X_2X^{2}_5X_7\}$, and for the second group
$\{Z^{2}_3X_6, X^{2}_3Z_6\}$. Depending on their measurement
outcomes, a two-qutrit pure entangled state, which is one of the
simultaneous eigenstates of $\{X^{2}_1X_4, Z_1Z_4\}$, is distilled
out between the parties $1$ and $4$.

Actually, one can verify that for any two parties $i \in
\{1,2,3,5,7\}$ and $j \in \{4,6\}$, a partition
$\{\{i,j\},T_2,T_3\}$ satisfying the condition (2) in Theorem 1
could be found, so $i$ and $j$ can share a two-qutrit pure
entangled state by forming the groups $T_2$ and $T_3$. For
example, for $i=2$ and $j=4$, such a partition is
$\{\{2,4\},\{1,3,7\},\{5,6\}\}$.

To our knowledge, this state is the first presented unlockable
bound entangled state in multiqutrit systems. Besides, by Theorem
2 we know that the Hilbert space of seven qutrits can be
decomposed into nine orthogonal subspaces such that the normalized
projection operator onto each of them is an unlockable bound
entangled state.

In similar manners, numerous unlockable bound entangled states in
arbitrary multiqudit systems can also be found. Moreover, one can
similarly use our lemma and theorems to analyze the properties of
these constructed states, such as symmetry under permutation of
parties, separability and unlockability, from the stabilizer
formalism.

\section{Extension to arbitrary multipartite systems}

In the previous sections, we considered only multiqudit systems.
Actually, the distillability and unlockability of the constructed
states $\rho_S$ depend mostly on the `local commutation' relation
of the stabilizer generators. The constraint that all parties
should have the same dimensions is really unnecessary. Our
definitions and theorems in Sec. II can be readily extended to
arbitrary multipartite systems.

More precisely, consider a $d_1 \times d_2 \times \dots \times
d_n$ system where the $i$th party has a $d_i$-dimensional space.
Define $G'(d_1,d_2,\dots,d_n)=\{g:g=\bigotimes_{i=1}^{n}{g_i}$
with $g_i=X^{a_i}_{(d_i)}$ or $Z^{b_i}_{(d_i)}$ for some $a_i,b_i$
$\}$. Then one can verify that for any element $g \in
G'(d_1,d_2,\dots,d_n)$, its eigenvalues are in the form
$\{1,\omega^{c},\omega^{2c},\dots,\omega^{D-c}\}$, where
$\omega=e^{i\frac{2\pi}{D}}$, $D$ is the least common multiple of
$d_1,d_2,\dots,d_n$, and $c|D$ .

Suppose we choose commuting elements $g_1,g_2,\dots,g_k$ from
$G'(d_1,d_2,\dots,d_n)$. Let $S=\langle g_1,g_2,\dots,g_k \rangle$
denote the Abelian group generated by them. Still we use $V_S$ to
denote the subspace stabilized by $S$. Then with the fact that
$\sum_{i=0}^{D-1}{\omega^{ci}}=0, \forall c=1,2,\dots,D-1$, one
can see that the projection operator onto $V_S$ is given by \begin{equation}
P_S=\prod_{i=1}^{k}{\frac{(I+g_i+g_i^2+\dots+g_i^{D-1})}{D}}. \end{equation}
And the maximally mixed state over $V_S$ is $\rho_S=P_S/tr(P_S)$.
Then following the same route of Sec.II.B, we can generalize the
three definitions and the Lemma 1, Theorem 1, Theorem 2 to the
elements in $G'(d_1,d_2,\dots,d_n)$.

Next we would like to use an example to illustrate this general
case. Consider a $2 \times 2 \times 4 \times 4 \times 6 \times 6$
system. Let \begin{equation} \begin{array}{l}
g_1=X_{(2)} \otimes Z_{(2)} \otimes X^2_{(4)} \otimes Z_{(4)} \otimes X^{3}_{(6)} \otimes Z_{(6)},\\
g_2=Z_{(2)} \otimes X_{(2)} \otimes Z_{(4)} \otimes X^2_{(4)} \otimes Z_{(6)} \otimes X^{3}_{(6)}.\\
\end{array} \end{equation} where $X_{(d)},Z_{(d)}$ are defined as
Eq.(\ref{equ:pauli}). $g_1$ and $g_2$ both have eigenvalues
$1,\omega,\omega^2,\dots,\omega^{11}$ where
$\omega=e^{i\frac{\pi}{6}}$. The maximally mixed state over the
subspace stabilized by $g_1,g_2$ is \begin{equation} \rho_{\langle g_1,g_2
\rangle}=\frac{1}{N}(\sum\limits_{i=0}^{11}{g_1^{i}})(\sum\limits_{j=0}^{11}{g_2^{j}}),
\end{equation} where $N=2\times 2 \times 4 \times 4 \times 6 \times 6$ is the
dimension of the whole space.

One can verify that $S=\langle g_1, g_2 \rangle $ is separable
with respect to any $2:2:2$ partition, e.g.
$\{\{1,2\},\{3,4\},\{5,6\}\}$. So this state is separable with
respect to any $2:2:2$ partition. In addition, any two parties can
obtain a pure entangled state by letting the other four parties
join together pairwise in an arbitrary fashion. This is because,
as one may check, any $2:2:2$ partition satisfies the condition
(2) in Theorem 1. For instance, consider the partition $\{\{1,6\},
\{2,3\},\{4,5\}\}$. Suppose the parties $2$ and $3$ join together,
and similarly for $4$ and $5$. If the group $\{2,3\}$ perform the
projective measurement in the basis of the simultaneous
eigenstates of $\{Z_{(2)}\otimes X^2_{(4)},X_{(2)}\otimes
Z_{(4)}\}$, and the group $\{4,5\}$ perform the projective
measurement in the basis of the simultaneous eigenstates of
$\{Z_{(4)}\otimes X^3_{(6)},X^2_{(4)}\otimes Z_{(6)}\}$, then
depending on their outcomes, a pure entangles state, which is a
simultaneous eigenstate of $\{X_{(2)}\otimes Z_{(6)}, Z_{(2)}
\otimes X^{3}_{(6)}\}$, will be obtained between $1$ and $6$.

It is worth noting that in this example although the six particles
have three different kinds of dimensions $2,4,6$, as shown above,
the unlockability of this state is very strong. So we learn that
the distinction between the dimensions of different parties is not
really an obstacle of building unlockable bound entangled states
in such systems. Nonetheless, we should point out that the
conditions in Theorem 1 may be not satisfiable for some
multipartite systems. One instance is the multipartite system in
which the dimensions of all parties are mutually relative prime.
But what we guarantee is that when the conditions in Theorem 1 are
fulfilled, we can use the theorem to build a class of unlockable
bound entangled states in the corresponding multipartite system.

\section{Conclusion}

In sum, we find an interesting relationship between two important
areas in quantum information science -- multipartite bound
entangled states and the stabilizer formalism. Our results provide
a simple way of constructing unlockable bound entangled states in
arbitrary multiqudit systems. These states not only can be
concisely described, but also possess properties which can be
easily explained from the stabilizer formalism. In particular, the
previous four-qubit Smolin states and its generalization to even
number of qubits can be viewed as special examples of our results.
Our theorems can also be extended to arbitrary multipartite
systems in which the dimensions of all parties may be different,
although their conditions may be in fact unsatisfiable in some
cases.

Finally, we would like to point out several directions for further
investigation along our way. The first one would be to extend our
work to more general situations. In our work we utilized the
inherent symmetry of Pauli matrices to construct our unlockable
bound entangled states. However, as the reader may have already
found out, our construction actually mainly relies on the `local
commutation' relation of the stabilizer generators. This relation
can be also defined over arbitrary multipartite operations which
can be written as the tensor products of unitary operations on
each subsystem, not just the generalized Pauli operations.
Therefore it is entirely possible that our definitions and
theorems can be appropriately adjusted so that it can be
applicable to a wider class of multipartite operations and states.
Another direction would be to study the properties and
applications of our constructed unlockable bound entangled stated,
such as their violation of Bell inequalities, whether they also
show the `superactivation' phenomenon, whether they can be used in
the information processing tasks such as remote information
concentration and multipartite key distribution. We hope that in
this way more interesting results about the structures and
features of multipartite bound entanglement will be found in the
future.

\section*{Acknowledgement}
We would like to thank Runyao Duan, Zhengfeng Ji, Yuan Feng and
Chi Zhang for helpful discussions. This work was partly supported
by the Natural Science Foundation of China (Grant Nos. 60621062
and 60503001) and the Hi-Tech Research and Development Program of
China (863 project) (Grant No. 2006AA01Z102).

\end{document}